\documentclass[twocolumn]{aastex63}
\received{June 1, 2019}
\revised{January 10, 2019}
\accepted{\today}
\submitjournal{ApJ}

\shorttitle{Boundary conditions in HD simulations with gravity}
\shortauthors{Caproni et al.}

\begin{document}

\title{Boundary conditions in hydrodynamic simulations of isolated galaxies and their impact on the gas-loss processes}

\correspondingauthor{Anderson Caproni}
\email{anderson.caproni@cruzeirodosul.edu.br}

\author[0000-0001-9707-3895]{Anderson Caproni}
\affiliation{N\'ucleo de Astrof\'\i sica, Universidade Cidade de S\~ao Paulo \\
	R. Galv\~ao Bueno 868, Liberdade, S\~ao Paulo, SP, 01506-000, Brazil}

\author{Gustavo A. Lanfranchi}
\affiliation{N\'ucleo de Astrof\'\i sica, Universidade Cidade de S\~ao Paulo \\
	R. Galv\~ao Bueno 868, Liberdade, S\~ao Paulo, SP, 01506-000, Brazil}

\author{Am\^ancio C. S. Fria\c ca}
\affiliation{Instituto de Astronomia, Geof\'\i sica e Ci\^encias Atmosf\'ericas, Universidade de S\~ao Paulo, R. do Mat\~ao 1226, Cidade Universit\'aria,\\
	S\~ao Paulo, SP, 05508-900, Brazil}

\author{Jennifer F. Soares}
\affiliation{N\'ucleo de Astrof\'\i sica, Universidade Cidade de S\~ao Paulo \\
	R. Galv\~ao Bueno 868, Liberdade, S\~ao Paulo, SP, 01506-000, Brazil}



\begin{abstract}

Three-dimensional hydrodynamic simulations are commonly used to study the evolution of the gaseous content in isolated galaxies, besides its connection with galactic star formation histories. Stellar winds, supernova blasts, and black hole feedback are mechanisms usually invoked to drive galactic outflows and decrease the initial galactic gas reservoir. However, any simulation imposes the need of choosing the limits of the simulated volume, which depends, for instance, on the size of the galaxy and the required numerical resolution, besides the available computational capability to perform it. In this work, we discuss the effects of boundary conditions on the evolution of the gas fraction in a small-sized galaxy (tidal radius of $\sim$1 kpc), like classical spheroidal galaxies in the Local Group. We found that open boundaries with sizes smaller than approximately 10 times the characteristic radius of the galactic dark-matter halo become unappropriated for this kind of simulation after $\sim$0.6 Gyr of evolution, since they act as an infinity reservoir of gas due to dark-matter gravity. We also tested two different boundary conditions that avoid gas accretion from numerical frontiers: closed and selective boundary conditions. Our results indicate that the later condition (that uses a velocity threshold criterion to open or close frontiers) is preferable since minimizes the number of reversed shocks due to closed boundaries. Although the strategy of putting computational frontiers as far as possible from the galaxy itself is always desirable, simulations with selective boundary condition can lead to similar results at lower computational costs.

\end{abstract}



\section{Introduction} \label{sec:intro}

Differential equations are widely used in different astrophysical contexts, from physical phenomena involving our solar system to the large-scale structures in the universe, as clusters and superclusters of galaxies. In particular, fluid-dynamic problems are formulated through differential equations that represent the conservation of mass, momentum, and energy of a fluid with a certain equation of state (e.g., \citealt{1987flme.book.....L}). 

To find a particular solution of any differential equation, it is necessary to provide some initial condition and/or some boundary conditions (BCs; e.g., \citealt{2005mmp..book.....A}). However, because of high complexity involving hydrodynamic (HD) problems in astrophysical systems, analytical solutions for the temporal behavior of a fluid are rare, leading to applications of numerical methods to solve the HD equations (e.g., \citealt{toro2009riemann}). There are several numerical codes dedicated to dealing with astrophysical gas/particle dynamics (e.g., \citealt{1992ApJS...80..753S, 2000ApJS..131..273F, 2000RMxAA..36...67R, 2002A&A...385..337T, 2003ApJ...589..444G, 2005ApJ...635..723A, 2005MNRAS.364.1105S, 2007ApJS..170..228M, 2014ApJS..211...19B}), adopting different strategies to numerically evolve gas/particle flows. Distinct BCs are usually available in those codes, which are chosen according to the specific physical situation to be studied.

In this work, we focus on HD simulations planned to study the time evolution of the gas content inside an isolated galaxy under the influence of a dark-matter distribution and supernova feedback (e.g., \citealt{1998MNRAS.299..249S, 1999ApJ...513..142M, 2003ApJ...590..778F, 2003ApJ...591...38W, 2006MNRAS.371..643M, 2007ApJ...667..170S, 2009A&A...501..189R, 2013MNRAS.429.1437R, 2014AdAst2014E...4R, 2015ApJ...805..109C, 2016ApJ...826..148E, 2017ApJ...838...99C, 2020arXiv200703702E}), aiming to verify the influence of the BCs on the gas removal efficiency.

This paper is structured as follows. In Section \ref{sec:selective_bound_cond}, we describe the three BCs analyzed in this work. Initial setup and the general results from the HD simulations performed in this work are presented in Section \ref{sec:num_simul} and discussed in Section \ref{sec:disc}. The main conclusions obtained in this work are
listed in Section \ref{sec:concl}.


\section{The selective boundary condition} \label{sec:selective_bound_cond}

Before introducing our selective boundary condition (SBC), it is useful to present the main characteristics of the additional boundary conditions (BCs) used in this work. 

Let $\rho$, $P$, and $\textbf{\emph{v}}$ be the mass density, the thermal pressure, and the velocity of a fluid element at a position $\textbf{\emph{r}}$ measured in a given reference frame. The former three quantities are usually referred to as primitive variables in HD problems. For grid numerical simulations, the region of interest is discretized on computational cells, where the HD equations are evolved in space and time. The region of interest, or simply the computational domain, is enclosed by numerical boundaries. Boundary conditions are implemented numerically by the usage of guard or ghost cells adjacent to the boundaries of the computational domain.

Let us also define $\hat{\textbf{\emph{n}}}$ as the unit vector orthogonal to the boundaries of a computational domain, always pointing outwards by convention. In the case of the open boundary condition (OBC), also known as outflow BC, the gradient of any primitive variable across the boundary along $\hat{\textbf{\emph{n}}}$ is set equal to zero (e.g., \citealt{2007ApJS..170..228M}).

\citet{2015ApJ...805..109C} and \citet{2017ApJ...838...99C} adopted closed boundary conditions (CBC) in their HD simulations. It differs from open boundaries only in terms of the values of $\textbf{\emph{v}}$ at the boundaries: all velocity components were set to zero in \citet{2015ApJ...805..109C}, while the null value was set only for the velocity component parallel to $\hat{\textbf{\emph{n}}}$, $\textbf{\emph{v}}_n=\textbf{\emph{v}}\cdot \hat{\textbf{\emph{n}}}$, in \citet{2017ApJ...838...99C}. Those authors adopted such boundary conditions to avoid that the frontiers of the computational domains in their simulations behaved as an infinity reservoir of matter due to the dark-matter gravitational potential (see Section \ref{sec:num_simul} for further discussion). A similar BC was also adopted by \citet{2003ApJ...590..778F}, where density and temperature at static boundaries ($\textbf{\emph{v}}=0$) is kept fixed at their initial values.

The SBC (also known as diode BC; e.g., \citealt{2000ApJS..131..273F, 2002ApJS..143..539Z}) is a variant of the CBC adopted in \citet{2017ApJ...838...99C}, in the sense that if the fluid element that reaches the boundary is moving outwards, as well as it having a speed higher than a predefined threshold value, $v_\mathrm{th}$, the CBC is switched to OBC at that location. In other words, the selective boundaries allow those fluids that are moving fast enough to leave the computational domain; otherwise SBC blocks their passage, keeping them inside the domain. Thus, the SBC can be defined as follows

\begin{eqnarray}
SBC & = & \left\{
\begin{array}{cc}
OBC, & \quad \text{if} \ \left(\textbf{\emph{v}}\cdot \hat{\textbf{\emph{n}}}>0 \ \text{and} \ \vert\textbf{\emph{v}}\cdot \hat{\textbf{\emph{n}}}\vert > v_\mathrm{th} \right) \\
CBC, & \text{otherwise} 
\end{array}\right.
\end{eqnarray}
where $SBC$ is the boundary condition to be used for a given position at the boundaries in a given time step, and $\vert\textbf{\emph{v}}\cdot \hat{\textbf{\emph{n}}}\vert$ $(=\vert\textbf{\emph{v}}_n\vert)$ is the the absolute value of $\textbf{\emph{v}}_n$ in a given cell adjacent to the boundary.

\begin{deluxetable*}{lc}
	\tablecaption{Some physical parameters of the isolated galaxy used in our simulations.\label{tab:Isogalaxy}}
	\tablewidth{0pt}
	\tablehead{
		\colhead{Parameter} & \colhead{Value} 
	}
	\startdata
	Dark-matter halo mass inside $R_{200}\tablenotemark{a}$  (M$_\sun$) & 3.1$\times 10^9$ \\
        $R_{200}$ (kpc) & 30.5\\
        Characteristic radius of the dark matter halo (kpc) & 0.3\\
        Maximum circular velocity due to dark matter (km s$^{-1}$) & 21.1\\
	Escape velocity at $R_{200}$ (km s$^{-1}$) & 64 \\
	Initial gas mass inside 950 pc\tablenotemark{b} (M$_\sun$) & 6.1$\times 10^8$ \\
	Initial gas temperature (K) & 5214 \\
	Initial gas number density\tablenotemark{c} (cm$^{-3}$) & 187 \\
	\enddata
	\tablenotetext{a}{The radius enclosing overdensity of 200 in relation to the critical density of the universe.}
	\tablenotetext{b}{The tidal radius of the galaxy.}
	\tablenotetext{c}{At the center of the galaxy.}
\end{deluxetable*}

\section{HD numerical simulations} \label{sec:num_simul}

\subsection{Initial setup} \label{subsec:ini_setup}

Aiming to test the impact of the boundaries on the evolution of the gas content inside an isolated (dwarf spheroidal) galaxy, we decided to use in our simulations a similar initial gas configuration found in \citet{2017ApJ...838...99C}. In a few words, an isothermal gas is put in hydrostatic equilibrium with a cored, static dark-matter gravitational potential (e.g., Equation 6 in \citealt{2017ApJ...838...99C}), so that its density distribution is peaked at the center of the gravitational potential well, decreasing radially as the galactocentric distance increases. 

Adopting a dwarf galaxy as a proxy for an isolated galaxy in our simulations avoids working with large computational domains, since dwarf galaxies are relatively small in size, with a tidal radius roughly below of some thousands of parsecs (e.g., \citealt{1998ARA&A..36..435M}). Consequently, it helps to conduct high numerical resolution experiments without a large number of computational cells, decreasing substantially the involved execution times.

We show in Table \ref{tab:Isogalaxy} the main physical characteristics of the isolated galaxy used in our numerical simulations. These values are compatible with those inferred for the classical dwarf spheroidal galaxy Ursa Minor.

\subsection{Perturbing the galactic gas: types Ia and II supernovae feedback} \label{subsec:SN}

The initial gas distribution is perturbed by supernova (SN) blasts in our simulations. We followed basically \citet{2017ApJ...838...99C} for the SN feedback recipe, even though the new version of our code used in this work follows independently types Ia and II supernovae\footnote{Further details of this new approach will be provided in a future paper in preparation \citep{Lanfranchi2022}.}. In a few words, the rates of types Ia and II SNe in our simulations were constrained by the chemical evolution model for Ursa Minor galaxy \citep{2004MNRAS.351.1338L, 2007A&A...468..927L}, a typical classical dwarf spheroidal galaxy in the Local Group. The imposed types Ia and II SNe rates are strictly respected during the whole of the simulations, telling to the code when an SN event must occur. On the other hand, where a SN event must take place depends on its type: denser regions are more prone to be selected for harboring a type II SN blast, while type Ia SNe are distributed randomly inside the galaxy. Independent of the type of supernova, an internal energy of $10^{51}$ erg is added into the computational cell elected as an SN site.

The SN feedback injects momentum into the interstellar medium, producing a net motion of the gas that is directed outward the galaxy. These galactic winds drive the gas losses that the simulated galaxy will experience as the time evolves. A portion of this galactic wind can reach the boundaries after a given interval, so that we must be concerned about the influence of the chosen boundaries on it.

\begin{deluxetable*}{cccccc}
	\tablecaption{Main characteristics of the three-dimensional HD simulations performed in this work.\label{tab:Simulations}}
	\tablewidth{0pt}
	\tablehead{
		\colhead{Name} & \colhead{$N_\mathrm{cell}$} & \colhead{$L$} & \colhead{$l$} & \colhead{$v_\mathrm{th}$} & \colhead{Uniform Grid?}\\
		\colhead{} & \colhead{} & \colhead{(kpc)} & \colhead{(pc cell$^{-1}$)} &\colhead{(km s$^{-1}$)} & \colhead{}
	}
	\decimalcolnumbers
	\startdata
	OBL3N100 & 100 & 3 & 30 & $-$\tablenotemark{a} & yes\\
 	OBL60N170 & 170 & 60 & $-$\tablenotemark{b} & $-$\tablenotemark{a} & no\tablenotemark{b}\\
	CBL3N100 & 100 & 3 & 30 & $-$\tablenotemark{c} & yes\\
	V64L12N200 & 200 & 12 & $-$\tablenotemark{d} & 64.0 & no\tablenotemark{d}\\
	V64L6N250 & 250 & 6 & $-$\tablenotemark{e} & 64.0 &  no\tablenotemark{e}\\
	V64L6N200 & 200 & 6 & 30 & 64.0 & yes\\
	V64L3N100 & 100 & 3 & 30 & 64.0 & yes\\
	V32L3N100 & 100 & 3 & 30 & 32.0 & yes\\
	V16L3N100 & 100 & 3 & 30 & 16.0 & yes\\
	V8L3N100 & 100 & 3 & 30 & 8.0 & yes\\
	V4L3N100 & 100 & 3 & 30 & 4.0 & yes\\
	V2L3N200 & 200 & 3 & 15 & 2.0 & yes\\
	V2L3N100 & 100 & 3 & 30 & 2.0 & yes\\
	V2L3N50 & 50 & 3 & 60 & 2.0 & yes\\
	V1.5L3N100 & 100 & 3 & 30 & 1.5 & yes\\
	V1L3N100 & 100 & 3 & 30 & 1.0 & yes\\
	\enddata
	\tablecomments{(1) Label of the simulation. (2) Total number of computational cells per Cartesian direction. (3) Linear size of the cubic computational domain in each Cartesian direction. (4) Numerical resolution. (5) Threshold value of the orthogonal velocity component at the boundaries used in the SBC simulations. (6) Whether the domain was divided uniformly in each direction.}
	\tablenotetext{a}{Derivatives of the primitive variables along orthogonal direction to the boundaries are set to the zero value.}
 	\tablenotetext{b}{In each Cartesian direction, we used 10 computational cells between -30 and -10.8 kpc and 10.8 and 30 kpc (1920 pc cell$^{-1}$), 5 cells between -10.8 and -6 kpc and 6 and 10.8 kpc (960 pc cell$^{-1}$), 5 cells between -6 and -3.6 kpc and 3.6 and 6 kpc (480 pc cell$^{-1}$), 5 cells between -3.6 and -2.4 kpc and 2.4 and 3.6 kpc (240 pc cell$^{-1}$), 5 cells between -2.4 and -1.8 kpc and 1.8 and 2.4 kpc (120 pc cell$^{-1}$), 5 cells between -1.8 and -1.5 kpc and 1.5 and 1.8 kpc (60 pc cell$^{-1}$), and 100 cells between -1.5 and 1.5 kpc (30 pc cell$^{-1}$).}
	\tablenotetext{c}{The normal velocity component at the boundaries is set to the zero value.}
	\tablenotetext{d}{For each Cartesian direction, we used 25 computational cells between -6.0 and -3.0 kpc and 3.0 and 6.0 kpc (120 pc cell$^{-1}$), 25 cells between -3.0 and -1.5 kpc and 1.5 and 3.0 kpc (60 pc cell$^{-1}$), and 100 cells between -1.5 and 1.5 kpc (30 pc cell$^{-1}$).}
	\tablenotetext{e}{In each Cartesian direction, we used 5 computational cells between -3.0 and -2.4 kpc and 2.4 and 3.0 kpc (120 pc cell$^{-1}$), 10 cells between -2.4 and -1.8 kpc and 1.8 and 2.4 kpc (60 pc cell$^{-1}$), 10 cells between -1.8 and -1.5 kpc and 1.5 and 1.8 kpc (30 pc cell$^{-1}$), and 200 cells between -1.5 and 1.5 kpc (15 pc cell$^{-1}$).}
\end{deluxetable*}

\subsection{Boundary conditions and instantaneous gas-loss rates} \label{subsec:boundcond_gasloss}

All the numerical HD simulations performed in this work made use of the PLUTO code\footnote{\url{http://plutocode.ph.unito.it/}} \citep{2007ApJS..170..228M} in its version 4.2. The classical hydrodynamic differential equations are evolved in time by a third-order Runge-Kutta algorithm, while the primitive variable reconstruction is done by a piecewise parabolic method (\citealt{1984JCoPh..54..174C}). The flux computation among numerical cells was done by the advection upstream splitting method (AUSM+; \citealt{1996JCoPh.129..364L}). We also assumed that gas respects the ideal equation of state, and it is under influence of a cored, dark-matter gravitational potential and cooling processes (see \citealt{2017ApJ...838...99C} for additional details).

We performed 16 HD numerical simulations to study the impact of the boundaries on the gas-loss rates. The main characteristics of these simulations are listed in Table \ref{tab:Simulations}. They include two simulations adopting OBC (OBL60N170 and OBL3N100), one simulation with CBC (CBL3N100), and the remaining 13 simulations used to test the behavior of the SBC.  

Except for the simulations OBL60N170, V64L12N200, V64L6N200, and V64L6N250, the computational domain consisted of a cubic box of size $L$ equals to 3 kpc. Numerical resolution in our simulations, $l$, was set from 15 to 60 pc per cell, which implied to a number of cells per axis, $N_\mathrm{cell}$, between 50 and 250. Except for OBL60N170, V64L12N200, and V64L12N250, all numerical experiments were performed using a uniform grid. All simulations were conducted in the Brazilian supercomputers SDumont \footnote{\url{http://sdumont.lncc.br}} and LAI \footnote{\url{lai.iag.usp.br}}. A total of $\sim$3.4$\times 10^6$ processor hours were required to run all the simulations presented in this work.

\begin{figure*}[ht!]
	\epsscale{0.8}
	\plotone{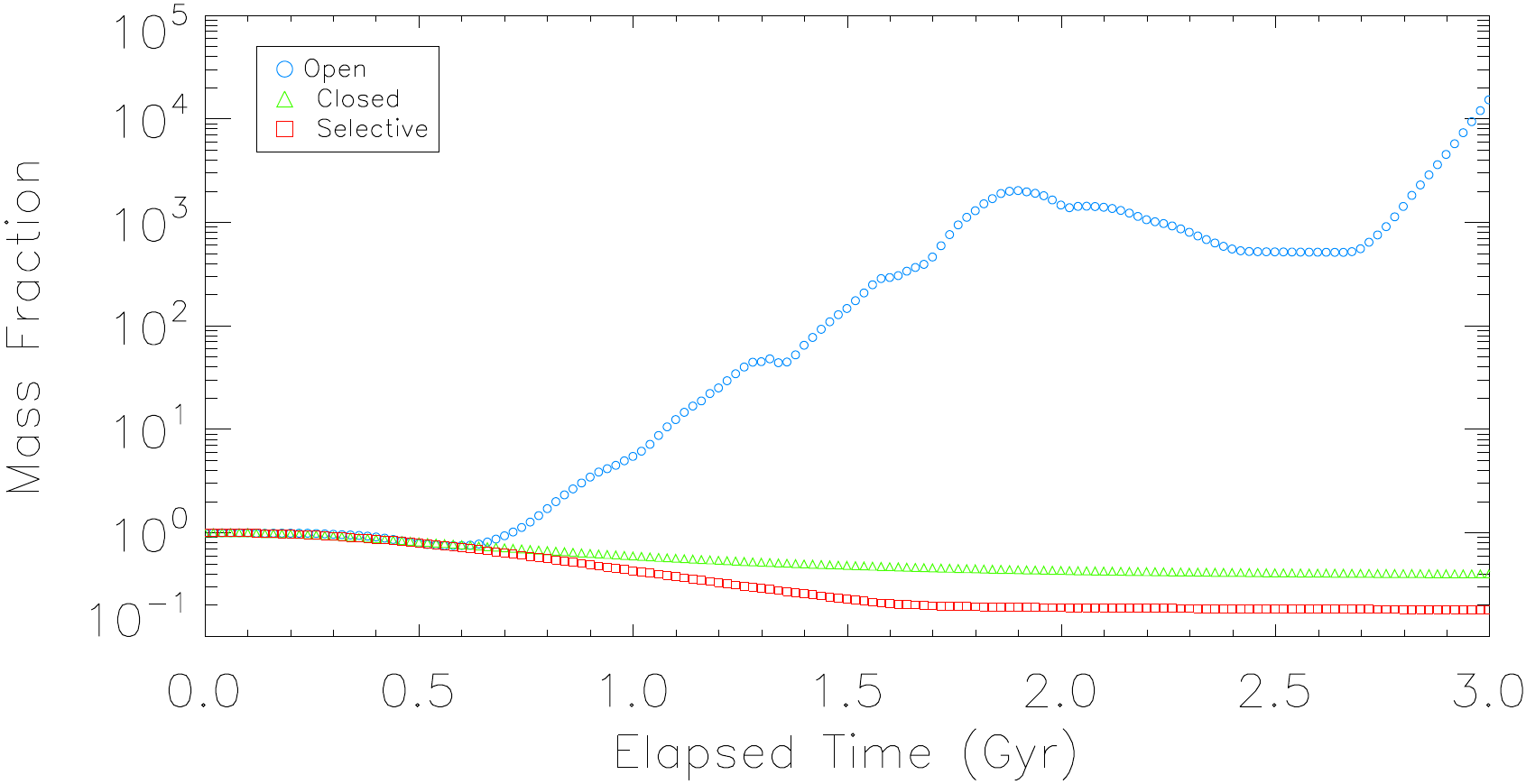}
	\caption{Instantaneous gas mass fraction inside a galactocentric radius of 950 pc (tidal radius of the galaxy) for the simulations using OBC (OBL3N100, blue circles), CBC (CBL3N100, green triangles), and SBC with $v_\mathrm{th} = 2$ km s$^{-1}$ (V2L3N100, red squares). These three simulations were made considering a cubic domain of $3^3$ kpc$^3$. \label{fig:open_close_sel}}
\end{figure*}

\begin{figure*}[ht!]
	\epsscale{0.8}
	\plotone{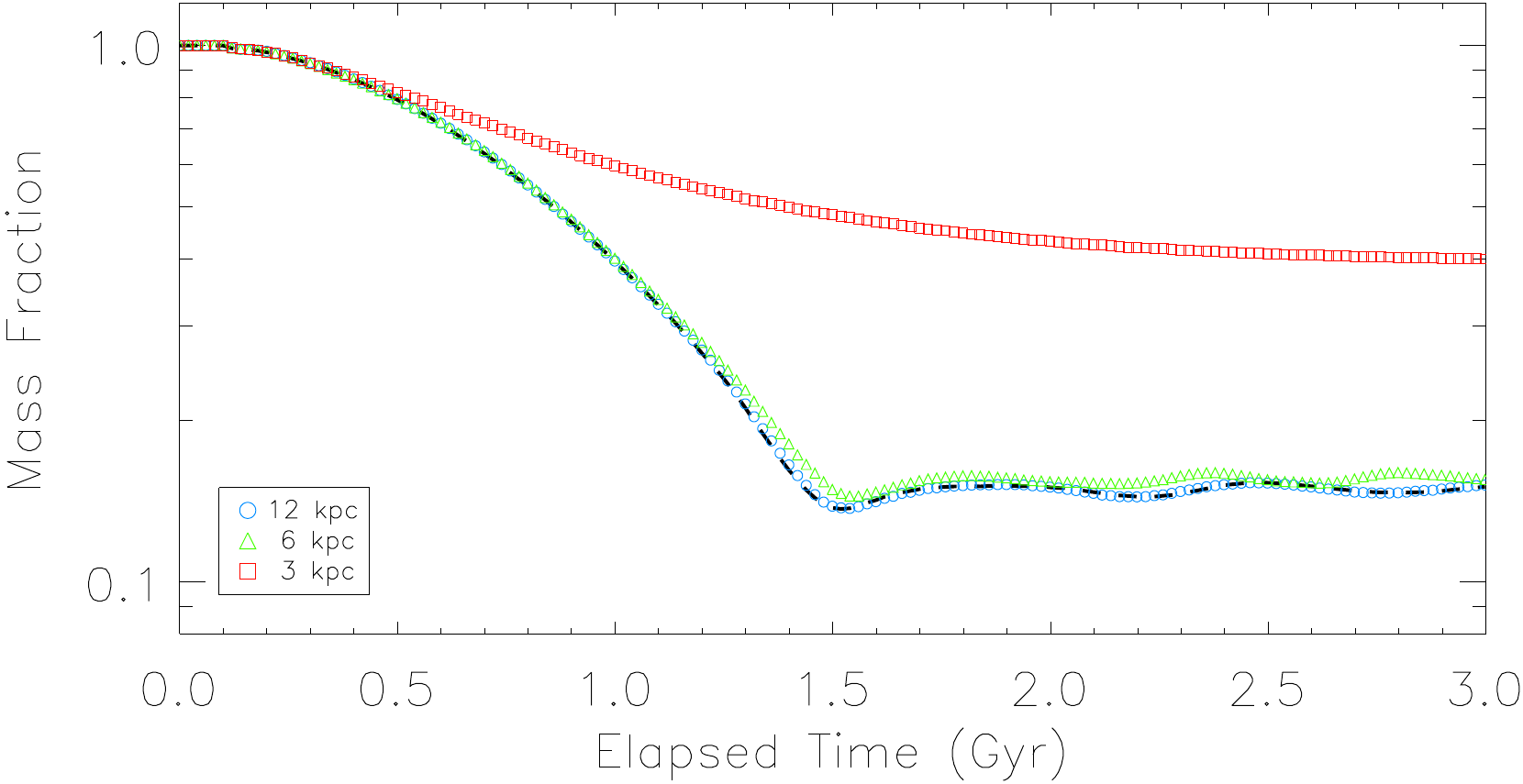}
	\caption{Instantaneous gas mass fraction inside a galactocentric radius of 950 pc (tidal radius of the galaxy) for SBC simulations with different sizes of the computational domain: $L = 12$ kpc (V64L12N200, blue circles), $L = 6$ kpc (V64L6N200, green triangles), and $L = 3$ kpc (V64L3N100, red squares). All these three runs adopt $v_\mathrm{th} = 64$ km s$^{-1}$. Dashed black line represents the results from an OBC simulation with $L = 60$ kpc (OBL60N170). \label{fig:dif_size_box}}
\end{figure*}

\subsubsection{Comparing open, closed and selective boundary condition simulations} \label{subsubsec:openclosedsel_boundcond}

Following \citep{2015ApJ...805..109C}, we estimated the instantaneous gas mass inside a galactocentric radius $R_\mathrm{gal} = 950$ pc (compatible with the tidal radius of the Ursa Minor dSph galaxy; e.g., \citealt{1995MNRAS.277.1354I}), after integrating numerically the mass density distribution obtained in the simulations

\begin{equation}\label{Inst_Gas_Mass}
M_\mathrm{gas}(t) = \int\int\limits_{V_\mathrm{gal}}\int\rho(x,y,z,t)dxdydz,
\end{equation}
where $M_\mathrm{gas}$ is the total gas mass at a time $t$ inside a spherical volume $V_\mathrm{gal}$ with a radius of $R_\mathrm{gal}$.

The instantaneous mass fraction of the gas inside $R_\mathrm{gal}$, $f_\mathrm{gas}$, is calculated from

\begin{equation}\label{Inst_Gas_Mass_fraction}
f_\mathrm{gas}(t) = \frac{M_\mathrm{gas}(t)}{M_\mathrm{gas,0}},
\end{equation}
where $M_\mathrm{gas,0}$ is the initial gas mass inside $R_\mathrm{gal}$ (see Table \ref{tab:Isogalaxy}). 

We show in Figure \ref{fig:open_close_sel} the behavior of $f_\mathrm{gas}$ for three distinct simulations: OBL3N100, CBL3N100, and V2L3N100 (see Table \ref{tab:Simulations} for further details). They show similar decreasing rates in the gas mass fraction due to SN feedback considering the first 600 Myr of evolution. After this interval, the situation changes dramatically: the OBC induces the breaking of the previous monotonic trend due to the rising of strong inflows of matter, which leads to extremely high (and nonphysical!) masses in comparison with what would be expected for a dwarf galaxy. This issue was already found by \citet{2015ApJ...805..109C} in their simulations: the OBC acts as an infinite reservoir of matter, which provides gas whenever the pressure equilibrium within the computational domain is broken due to the domain discretization (e.g., \citealt{2002ApJS..143..539Z}).

On the other hand, the decrease of the amount of gas still remains after 600 Myr for both CBL3N100 and V2L3N100 runs, even though the loss rates lower gradually until they become almost null after $\sim$1.6 Gyr. There is also a systematic offset between the gas mass fractions from CBL3N100 and V2L3N100 runs, the former presenting a substantially higher value after an elapsed time of 3 Gyr ($\sim$0.40\footnote{This value is in good agreement with that found by \citet{2017ApJ...838...99C} using a higher numerical resolution ($\sim$12 pc) in comparison with the simulations presented in this work.} against $\sim$0.18 for V2L3N100). The reason is that gas flows reaching boundaries with speeds higher than $v_\mathrm{th} = 2$ km s$^{-1}$ are allowed to leave the computational domain, while the CBC retains them, independent of how fast the gas flow is.

The quasi-saturation in the gas removal is detected in both CBL3N100 and V2L3N100 runs. The same result was also found by \citet{2017ApJ...838...99C} (see their Figure 4). \citet{2017ApJ...838...99C} also pointed out that reverse shocks generated when the SN-driven gas reaches the computational boundaries could decrease the inferred gas losses due to the gas retention. From a simple analytical calculation using the escape velocity associated with the dark-matter (DM) halo, these authors estimated that the CBC decreased the gas losses by a factor of $\sim$2.5 after 3 Gyr of evolution.

\begin{figure*}[ht!]
	\epsscale{0.8}
	\plotone{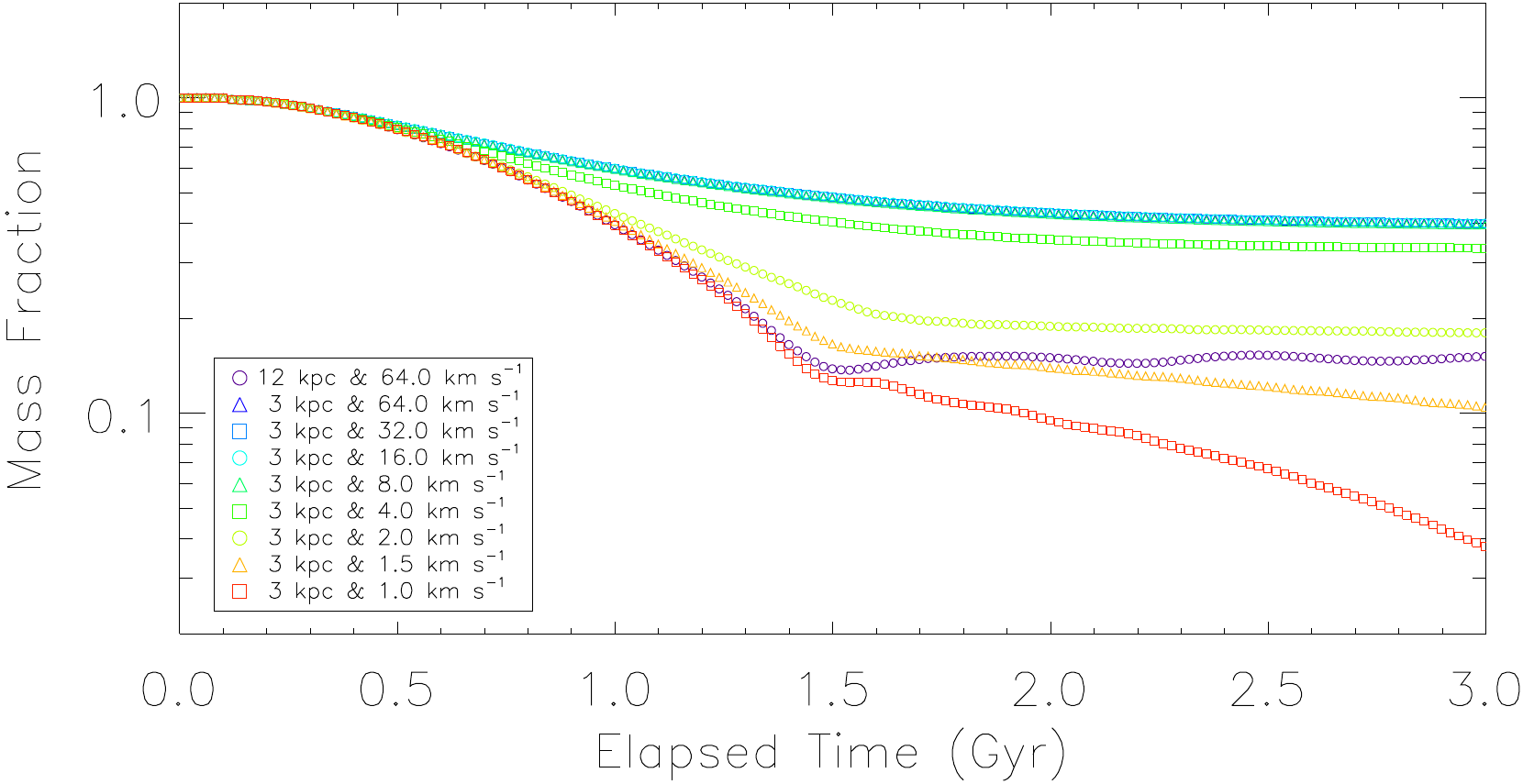}
	\caption{Instantaneous gas mass fraction inside a galactocentric radius of 950 pc (tidal radius of the galaxy) for SBC simulations with different threshold speeds (from 64 to 1 km s$^{-1}$). Simulation V64L12N200 is also plotted (purple circles).  \label{fig:thresh_vel}}
\end{figure*}

An alternative way to estimate the influence of the computational frontiers on the results is to put them farther and compare the amount of gas left inside the galaxy after a certain elapsed time. Thus, we run two additional simulations, V64L6N200 and V64L12N200, in which the computational domain was extended, respectively, by a factor of 2 and 4 in relation to the original box size. These two simulations were compared with the V64L3N100 run, in which $v_\mathrm{th}$ was set to 64 km s$^{-1}$ (the DM escape velocity used by \citet{2017ApJ...838...99C} in their analytical calculations). The time evolution of the gas mass fraction inside $R_\mathrm{gal}$ for these three simulations is shown in Figure \ref{fig:dif_size_box}. 

Until $\sim$500 Myr, these three simulations produced the same instantaneous gas mass fractions, as it can be seen in Figure \ref{fig:dif_size_box}. After that, the discrepancy between V64L3N100 and the other two runs increases monotonically until approximately 1.5 Gyr, when such differences stabilize approximately around a factor of about 2.7. Note this factor is compatible with the value estimated previously by \citet{2017ApJ...838...99C} from arguments based on the comparison between the escape velocity of the dark-matter halo and the velocities in the adjacent cells at the computational boundaries.

Simulations V64L6N200 and V64L12N200 agree with each other, indicating that a computational domain with a size of 6 kpc ($\sim$6 times the tidal radius of our fiducial galaxy) is enough to minimize boundary effects on the gas losses in similar simulations performed in this work. Besides, a low-amplitude oscillatory behavior in the curves concerning these two simulations is seen in Figure \ref{fig:dif_size_box}. We have interpreted this as the result of the competition between outward galactic winds driven by SNe and the DM halo's gravity, which tries to push gas back into the galaxy. In other words, these oscillations could be the numerical realization of the ``keeping the gas spread and heated" suggested previously in the literature (e.g., \citealt{2006MNRAS.371..885R, 2017ApJ...838...99C}). 

Even though Figure \ref{fig:dif_size_box} indicates some convergence in the gas mass fraction derived from the simulations V64L6N200 and V64L12N200, their domain dimensions are small in comparison to $R_\mathrm{200}$ ($\sim$30 kpc; see Table \ref{tab:Isogalaxy}) of the dark-matter halo of our fiducial galaxy. To verify whether these results are indeed representative of the gas losses driven by supernovae, we run an additional OBC simulation, OBL60N170, with a length of 60 kpc ($\sim$$2R_\mathrm{200}$) in each Cartesian direction. Its derived instantaneous mass fraction is shown by a dashed black line in Figure \ref{fig:dif_size_box}, indicating a quite similar behavior inferred from the simulations V64L6N200 and V64L12N200. Thus, a domain size of about 6 kpc seems to be enough for HD simulations of isolated galaxies with similar properties listed in Table \ref{tab:Isogalaxy}.

\begin{figure*}[ht!]
	\epsscale{0.8}
	\plotone{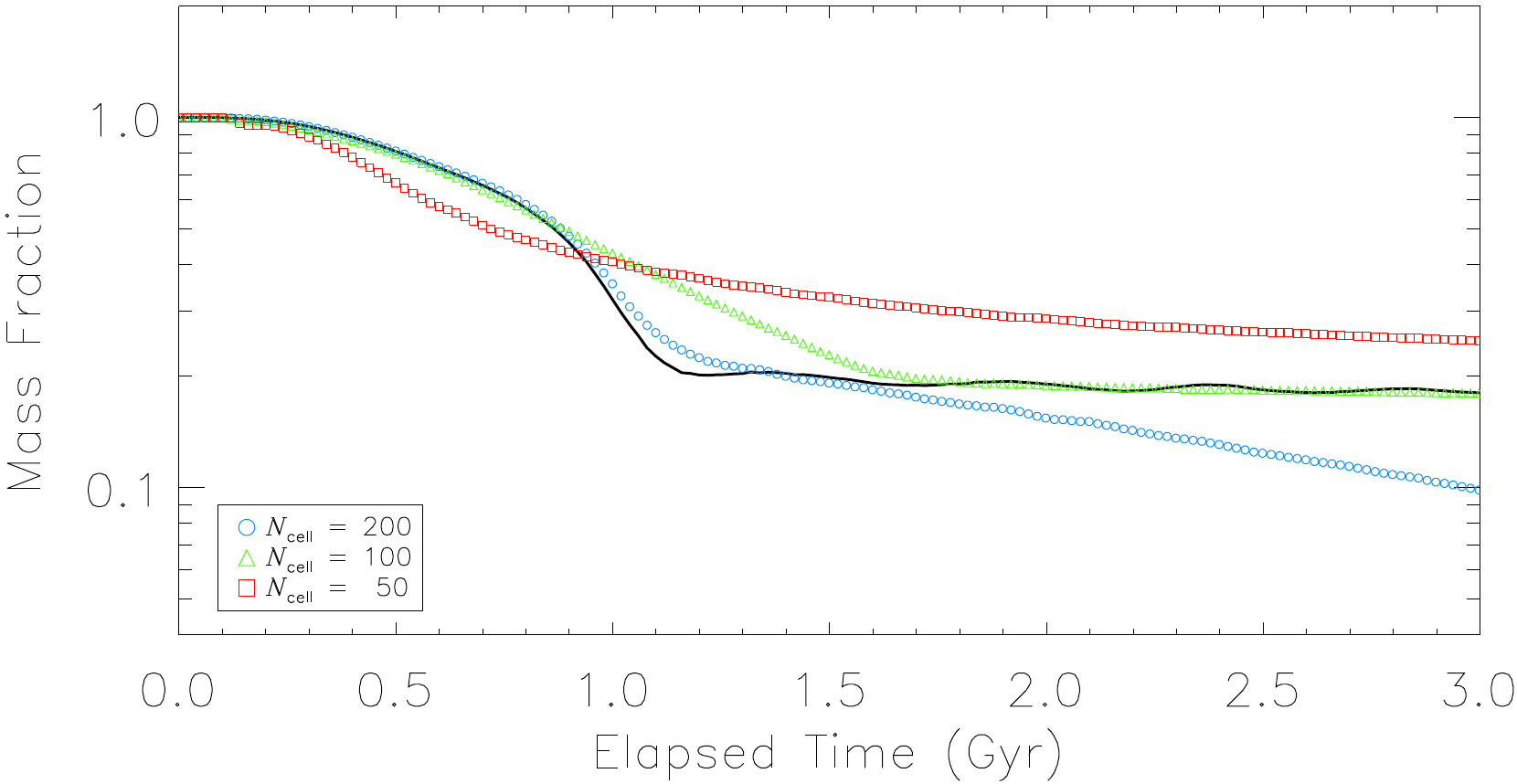}
	\caption{Instantaneous gas mass fraction inside a galactocentric radius of 950 pc for the SBC simulations V2L3N200 (blue circles), V2L3N100 (green triangles), and V2L3N50 (red squares). All these simulations use the same value for $v_\mathrm{th}$ (= 2 km s$^{-1}$), but differing in terms of numerical resolution. The solid black line shows the results from the simulation V64L6N250. \label{fig:num_res}}
\end{figure*}

\subsubsection{The impact of the threshold velocity on the selective boundary condition simulations} \label{subsubsec:threshold_velocity}

As it was discussed in the previous section, the usage of $v_\mathrm{th} = 64$ km s$^{-1}$ in SBC simulation V64L3N100 increased the amount of gas left after 3 Gyr of evolution by a factor of 2.7 in comparison to the simulations V64L6N200 and V64L12N200, which made use of larger computational domains. A question that arises is whether it is possible to recover the results found in larger box simulations just varying the value of $v_\mathrm{th}$ in the SBC simulations. Thus, we run seven additional simulations with the same initial setup and resolution of V64L3N100, but decreasing the value of $v_\mathrm{th}$ mostly by multiples of 2. A comparison among these simulations, as well with the larger box simulation V64L12N200 can be seen in Figure \ref{fig:thresh_vel}.

Again, no apparent differences among all simulations in Figure \ref{fig:thresh_vel} are noted until approximately 500 Myr of evolution. After this interval, the instantaneous amount of gas left inside $R_\mathrm{gal}$ decreases systematically as $v_\mathrm{th}$ is lowered from 64 to 1 km s$^{-1}$. The largest differences occur when $v_\mathrm{th} \la 4$ km s$^{-1}$, indicating that only a small portion of the gas that is pushed away by the SNe reaches the boundaries with speeds higher than $\sim$4 km s$^{-1}$. 

It can be seen in Figure \ref{fig:thresh_vel} that simulations V2L3N100 and V1.5L3N100 led to gas mass fractions closer to that obtained in the simulation V64L12N200, indicating that the appropriated value of $v_\mathrm{th}$ must be roughly between 1.5 km s$^{-1}$ and 2.0 km s$^{-1}$ for simulations with a cubic domain of 3$\times$3$\times$3 kpc$^3$.

\subsubsection{Effects of the numerical resolution} \label{subsubsec:num_resolution}

Adopting simulation V2L3N100 as a reference, we multiplied (divided) by a factor of 2 the number of computational cells, but keeping all additional parameters fixed, generating simulation V2L3N200 (V2L3N50). It means a change in the numerical resolution from 30 pc cell$^{-1}$ to 15 pc cell$^{-1}$ in the case of V2L3N200, while a resolution of 60 pc cell$^{-1}$ is attained for the simulation V2L3N50. We show in Figure \ref{fig:num_res} the influence of the numerical resolution on the instantaneous amount of gas left inside the galaxy. No difference in the mass fraction among these three simulations is seen during the first 200 Myr, when V2L3N50 begins to show a higher mass-loss rate in comparison to the other ones. This trend is inverted after about 1 Gyr and remains so until the end of the simulations.

Concerning simulations V2L3N100 and V2L3N200, there is no significant difference between them until $\sim$1 Gyr, but after 1.5 Gyr, $f_\mathrm{gas}$ decreases slowly in V2L3N200, in contrast with simulation V2L3N100 that presents a small-amplitude oscillations around $f_\mathrm{gas}\sim 0.185$. The increment in the numerical resolution from 30 to 15 pc cell$^{-1}$ is allowed to solve the snowplow transition radius for number densities as low as 1 cm$^{-3}$ (e.g., \citealt{1988ApJ...334..252C, 1988RvMP...60....1O}), avoiding over cooling issues that weaken the kinetic feedback from supernovae (e.g., \citealt{2011MNRAS.415.3706C, 2015ApJ...809...69S, 2017ApJ...838...99C}). Thus, a larger fraction of gas reached speeds higher than the threshold speed of 2 km s$^{-1}$ in simulation V2L3N200, leaving definitively the computational domain. 

At this point, it is interesting to verify whether the monotonic decrease of $f_\mathrm{gas}$ in V2L3N200 is due to a rather low value of $v_\mathrm{th}$ in SBC. For this aim, we run an additional simulation, V64L6N250, where we doubled the size of the computational domain but keeping the numerical resolution of 15 pc per cell inside a cubic subdomain of 3 kpc in size (see Table \ref{tab:Simulations} for further details). The behavior of $f_\mathrm{gas}$ as a function of time is shown in Figure \ref{fig:num_res} by the black solid line. No monotonic decrease of $f_\mathrm{gas}$ after 1.5 Gyr is seen but there are small-amplitude variations around $f_\mathrm{gas}\sim 0.185$ instead, as in simulation V2L3N100. It suggests that $v_\mathrm{th} = 2$ km s$^{-1}$ adopted in V2L3N200 is subestimated somehow. Based on the results shown in Figure \ref{fig:thresh_vel}, an increment of about 0.5-1.0 km s$^{-1}$ in $v_\mathrm{th}$ might be enough to reconcile simulations V2L3N200 and V64L6N250.
 
Finally, we can also note that the increment of numerical resolution led to a low amount of gas left inside the galaxy after 3 Gyr of evolution, a factor of $\sim$$2.5$ between the lowest and highest numerical simulations in Figure \ref{fig:num_res} (V2L3N50 and V2L3N200, respectively). This difference is not too big if we consider the usual uncertainties regarding the estimates of the mass in stars, gas and dark matter in galaxies, as well as a relatively poor knowledge concerning the individual efficiencies of the feedback mechanisms to remove gas in those systems. As it was already mentioned, a small fine tuning in the value of $v_\mathrm{th}$ can diminish or even eliminate those discrepancies.

\begin{figure*}[ht!]
	\epsscale{2.2}
	\plottwo{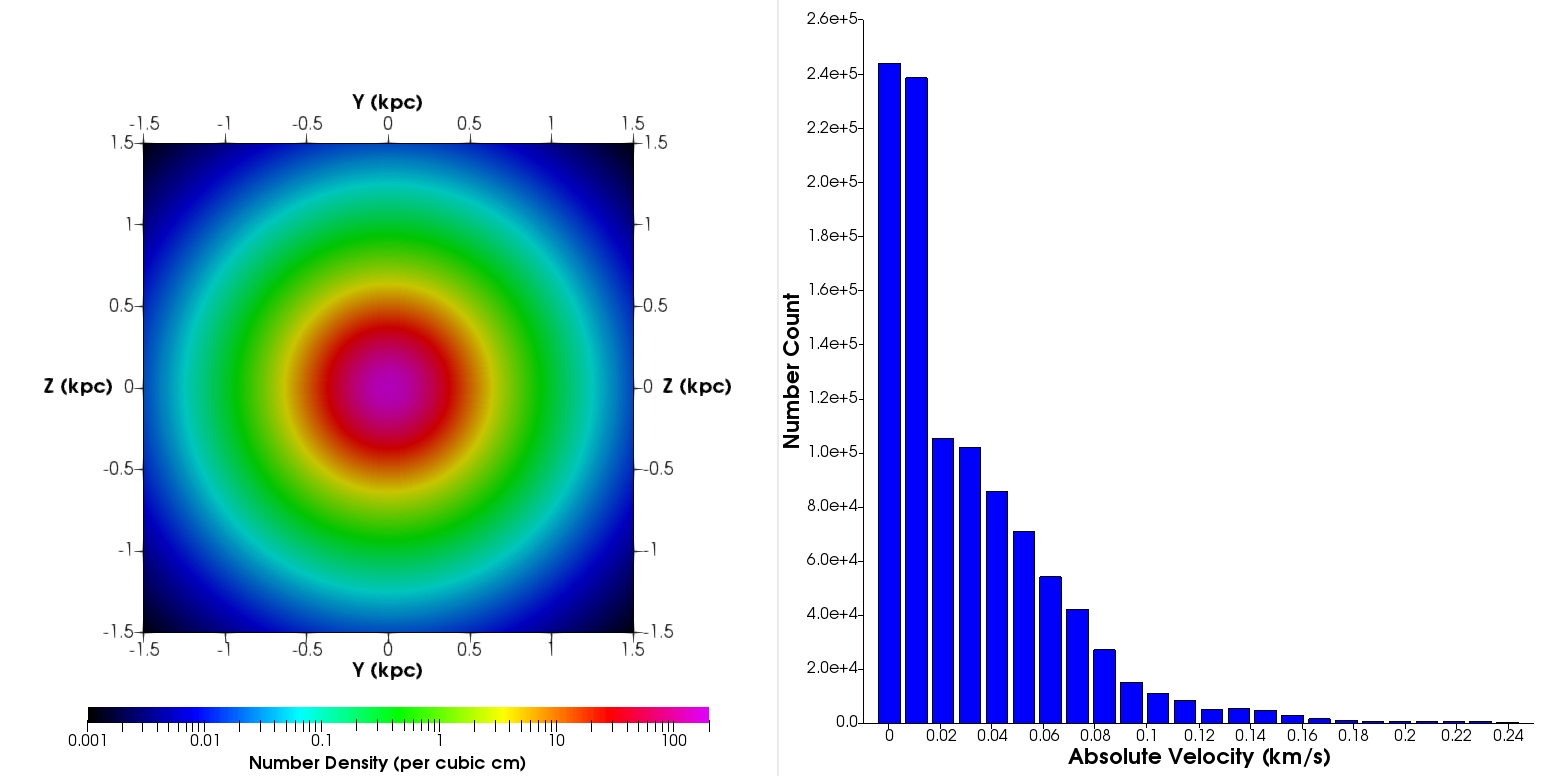}{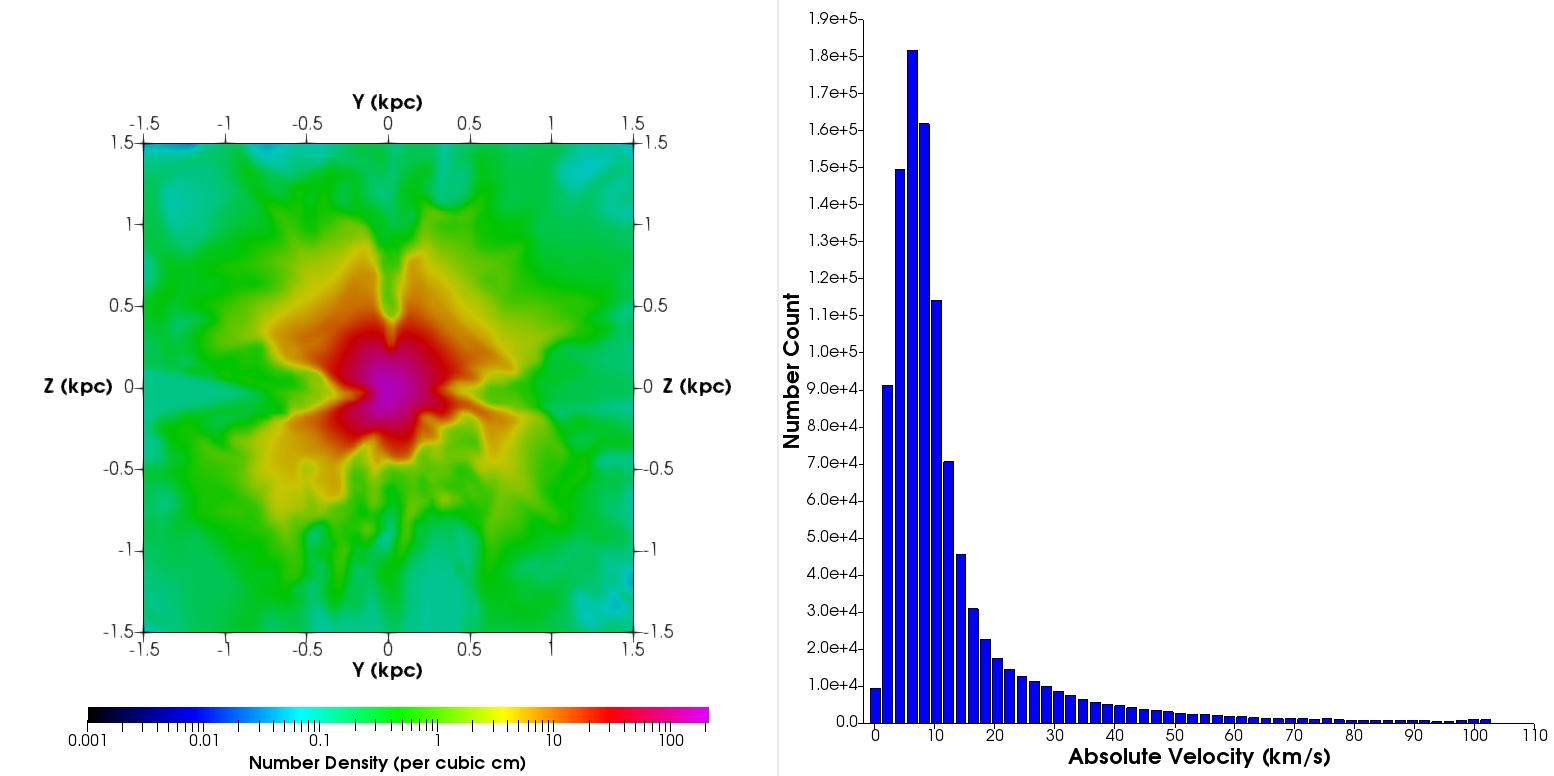}
	\caption{Effects of the SBC (upper panels) and OBC (bottom panels) on the stability of the initial gas configuration under hydrostatic equilibrium with a cored DM gravitational potential after 500 Myr of evolution and considering a cubic domain of 3 kpc$\times$3 kpc$\times$3 kpc. Left panels display the number density distribution on $X=0$ plane, while the right panels show the histogram of the absolute velocity of the gas in the whole computational domain for both BCs. \label{fig:stability}}
\end{figure*}

\section{Discussion} \label{sec:disc}

Our results showed that no influence of the BCs on the instantaneous gas-loss rates is observed until $\sim$600 Myr in the simulations discussed in Section \ref{subsec:boundcond_gasloss}. It suggests that noncosmological grid-based HD simulations involving isolated galaxies will not be substantially influenced by the choice of a particular BC if the simulated time is less or of the order of some hundreds of Myr, as it is the case of several previous works involving different types of galaxies (e.g., \citealt{1999ApJ...513..142M,2003ApJ...590..778F,2003ApJ...591...38W,2004ApJ...617.1077F,2015MNRAS.446..299M,2015ApJ...812...90M,2019MNRAS.482.1304E,2020arXiv200703702E}). For analogous HD simulations but involving longer timescales of evolution, the usage of SBC may be a useful alternative without sacrificing the numerical resolution and/or increasing the computational costs in the case of putting the numerical frontiers far from the galaxy (e.g., \citealt{2006MNRAS.371..643M,2000ApJ...538..559M,2016ApJ...826..148E}). To provide a sense of the gain in CPU time, we run four additional simulations evolved during 200 Myr in a workstation equipped with 128 2.2 GHz processors. Two of these simulations adopt SBC with $v_\mathrm{th} = 2$ km s$^{-1}$, while the two complementary ones use OBC. Besides, domain volumes of $3\times3\times3$ and $60\times60\times60$ kpc$^3$ were built for both SBC and OBC. In the case of the small volume domain ($3\times3\times3$ kpc$^3$), 60 computational cells per Cartesian axis were generated, implying a numerical resolution of 50 pc per cell. For the larger volume simulations, we kept the same numerical resolution of 50 pc per cell between -1.5 and 1.5 kpc, but decreasing nonmonotonically this resolution until it reaches the numerical boundaries at -30 and 30 kpc, leading to 102 cells per Cartesian direction. The results can be summarized as follows:

\begin{itemize}

\item The size of the computational domain is fixes, and the elapsed time to complete a simulation does not depend strongly on the assumed BC: in the case of a domain size of 3 kpc, $\sim$12.8 and 13.1 hours for SBC and OBC, respectively; for a 60-kpc box, the elapsed times were $\sim$48.7 and 48.6 hr for SBC and OBC, respectively;

\item However, a larger domain implies in a substantial longer time for the completion of the simulation. For instance, a larger computational domain with OBC led to a longer execution time by a factor of $\sim$3.7. Even though this factor is smaller than the ratio between the total number of the cells used in the simulations, $(102/60)^3\sim 4.9$, it shows that larger domains imply higher computational costs that could become prohibitive if high-performance computational resources are not accessible in practice. 

\end{itemize}

Besides the avoidance of the frontiers of the computational domain behaving as an infinity reservoir of matter in simulations with gravity, the SBC can decrease (or even eliminate) the occurrence of reversing flows of matter due to a pure CBC. As any flow colliding with a CBC will have its normal velocity reversed, it induces spurious backflows of matter that could modify the previous gas motions at interacting zones, as well as the physical conditions (density and temperature) of the gas (mainly if the created backflows become strong shocks). Note also that these reversing flows are expected to occur even in the absence of gravity forces.

The choice of a particular BC may also influence on the stability of the initial gas configuration under hydrostatic equilibrium with a gravitational potential. To quantify this effect on grid-based simulations of isolated galaxies, we rerun simulations OBL3N100 and V2L3N100 for 500 Myr turning off the SN feedback during the whole simulation. The results are shown in Figure \ref{fig:stability}. We note in the case of SBC (upper panels in Figure \ref{fig:stability}) that the initial gas distribution is well preserved during the whole simulation, with spurious speeds being lower than 0.25 km s$^{-1}$ ($\sim$50 percent of the cells have speeds lower than 20 m s$^{-1}$). On the other hand, the usage of OBC in a relatively small computational domain (lower panels in Figure \ref{fig:stability}) induces catastrophic inflows of gas that destroyed the initial spherically symmetric cored distribution of the gas, producing spurious speeds as high as some tens of kilometers per second. To reduce such spurious motions using OBC, larger computational domains are needed.

\begin{figure}[ht!]
	\epsscale{1.18}
	\plotone{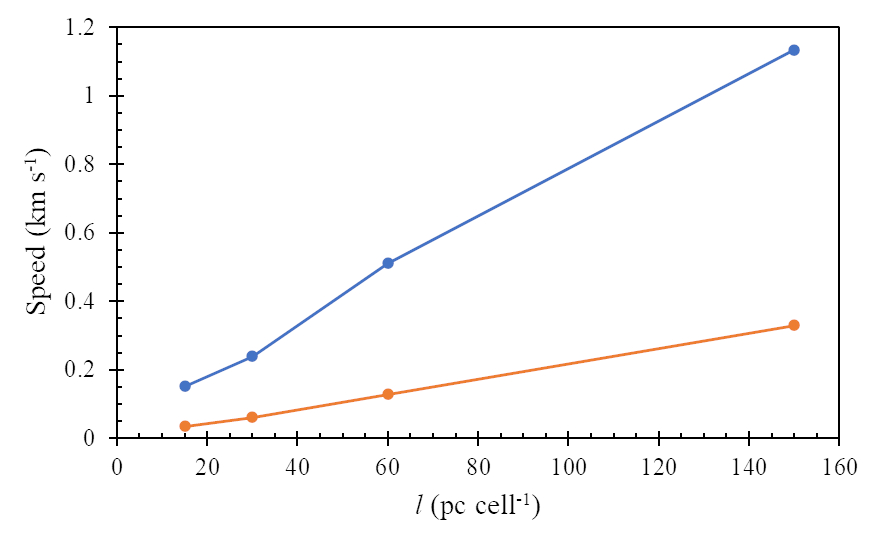}
	\caption{Maximum and mean spurious speeds (blue and orange curves, respectively) as a function of the numerical resolution of non-SN feedback simulations with SBC after 500 Myr of evolution.  \label{fig:stability_numres}}
\end{figure}

The magnitude of the spurious accretion also depends on the numerical resolution, as pointed out previously by \citet{2002ApJS..143..539Z}. We analyzed the impact of the numerical resolution on the time stability of the initial gas configuration rerunning simulations V2L3N50, V2L3N100, V2L3N200 without SN feedback, including also an extra simulation with a lower numerical resolution in comparison with the previous ones ($l = 150$ pc cell$^{-1}$). We show in Figure \ref{fig:stability_numres} the behavior of the spurious speeds in terms of numerical resolution after 500 Myr of evolution considering SBC. Trends of the increase of the mean and maximum spurious speed with the decrease of the numerical resolution are clearly seen in Figure \ref{fig:stability_numres}, even though their values are always very small ($\la 1$ km s$^{-1}$) in comparison to the OBC simulation shown in  Figure \ref{fig:stability}. These results suggest that SBC may be also useful in numerical problems somehow involving the hydrostatic equilibrium condition.

\section{Conclusions} \label{sec:concl}

In this work, we studied the influence of the computational frontiers on the gas removal process in (small) galaxies. The option for using an initial configuration compatible with a typical dwarf galaxy (tidal radius of about 1 kpc) is justified by keeping the computational domain as small as possible without sacrificing substantially the numerical resolution, and keeping the computational costs relatively low as well. Three different boundary conditions were employed in this work: open (or outflow), closed, and selective boundary conditions. The 16 hydrodynamic simulations with types Ia and II supernovae feedback performed in this work adopted a cubic domain where the galactic center coincides with the center of the computational box. The majority of these simulations have frontiers put at a galactocentric distance corresponding to $\sim$ 1.6$R_\mathrm{gal}$. Our main results are summarized as follows.

\begin{itemize}
	
	\item No difference in the gas mass fraction left inside the galaxy is noted until about 600 Myr of evolution, independent of the three boundary conditions analyzed in this work. It suggests that similar simulations involving short periods of time can adopt open boundary conditions without any loss of integrity of the results;
	
	\item After 600 Myr of evolution, open boundary conditions for a relatively small computational box (sizes smaller than $\sim3R_\mathrm{gal}$ or about 10 times the characteristic radius of the galactic dark-matter halo) act as an infinity reservoir of gas due to dark-matter gravity whenever the pressure equilibrium within the computational domain is broken due to the domain discretization (e.g., \citealt{2002ApJS..143..539Z}). In this case, closed or selective boundary conditions are preferable if the increase of the computational edges are somehow unfeasible;

	\item As it was already expected (e.g, \citealt{2015ApJ...805..109C}), closed frontiers tend to retain more gas in comparison to the selective boundary condition, impacting on the amount of mass left inside the galaxy: a factor of 2 approximately (see Figure \ref{fig:open_close_sel});

	\item Concerning the influence of the value of $v_\mathrm{th}$ used in the selective boundary condition simulations, no difference in $f_\mathrm{gas}$ is seen until approximately 500 Myr of evolution. It remains true until 3 Gyr for the simulations using $v_\mathrm{th}\ga 8$ km s$^{-1}$, coinciding with the results from the closed boundary simulation. For the simulations with $v_\mathrm{th}\la 4$ km s$^{-1}$, the instantaneous amount of gas left inside the galaxy decreases systematically as $v_\mathrm{th}$ is lowered;

	\item For $v_\mathrm{th}\la 1.5$ km s$^{-1}$, $f_\mathrm{gas}$ decreases with time, in contrast with SBC simulations with higher $v_\mathrm{th}$ that present a plateau-like behavior after $\sim$1.5 Gyr of evolution. Numerical simulations with larger computational domains ($\ga 6R_\mathrm{gal}$) show similar plateau-like behavior, but showing also a small-amplitude oscillation around  $f_\mathrm{gas}\sim 0.185$ possibly produced by the competition between the pull from the dark-matter gravitational potential and the push due to the supernova feedback;
	
	\item In terms of numerical resolution, our results show no difference in the mass fraction during the first 200 Myr when $l$ is varied from 60 to 15 pc cell$^{-1}$. This interval is extended to about 1 Gyr considering simulations with 30 and 15 pc cell$^{-1}$ only. The monotonic decreasing of $f_\mathrm{gas}$ seen in V2L3N200 is not present in V64L6N250 with a larger computational domain,  indicating that $v_\mathrm{th} = 2$ km s$^{-1}$ adopted in V2L3N200 is subestimated somehow. Based on the results shown in Figure \ref{fig:thresh_vel}, a small increment of about 0.5-1.0 km s$^{-1}$ in $v_\mathrm{th}$ might be enough to reconcile simulations V2L3N200 and V64L6N250.
	
	\item Although the strategy of putting computational frontiers as far as possible from the galaxy is always desirable, our simulations with a selective boundary condition can lead to similar results but at less expensive demands regarding computational resources.
	
\end{itemize}

As a final remark, even though we have analyzed the influence of the boundary conditions over the gas-loss rates using a dwarf spheroidal galaxy, the SBC strategy can be adopted for any type of galaxy or astrophysical system that demands closed numerical frontiers (e.g., see \citealt{2021ApJ...914...32L} for an application involving SBC in the context of intermediate-mass black hole feedback in dwarf spheroidal galaxies).


\acknowledgments

A.C., G.A.L., and J.F.S. thank the Brazilian agency FAPESP (grants 2014/11156-4, 2017/25651-5, 2017/25799‐2, 2019/21615‐0, and 2022/16883‐8). The authors acknowledge the National Laboratory for Scientific Computing (LNCC/MCTI, Brazil) for providing HPC resources of the SDumont supercomputer (\url{http://sdumont.lncc.br}), which have contributed to the research results reported within this paper. This work has made use of the computing facilities of the Laboratory of Astroinformatics (IAG/USP, NAT/UCS), whose purchase was made possible by the Brazilian agency FAPESP (grant 2009/54006-4) and the INCT-A. We would like to thank the anonymous referee for the constructive report.

%




\software{ParaView, \citep{paraview2005, paraview2015}, IDL (Interactive Data Language, \url{https://www.harrisgeospatial.com/Software-Technology/IDL})}




\bibliography{bib}

\begin{thebibliography}{}
\expandafter\ifx\csname natexlab\endcsname\relax\def\natexlab#1{#1}\fi
\providecommand{\url}[1]{\href{#1}{#1}}
\providecommand{\dodoi}[1]{doi:~\href{http://doi.org/#1}{\nolinkurl{#1}}}
\providecommand{\doeprint}[1]{\href{http://ascl.net/#1}{\nolinkurl{http://ascl.net/#1}}}
\providecommand{\doarXiv}[1]{\href{https://arxiv.org/abs/#1}{\nolinkurl{https://arxiv.org/abs/#1}}}

\bibitem[{{Ahrens} {et~al.}(2005){Ahrens}, {Geveci}, \& {Law}}]{paraview2005}
{Ahrens}, J., {Geveci}, B., \& {Law}, C. 2005, {ParaView: An End-User Tool for
  Large Data Visualization}, 17

\bibitem[{{Anninos} {et~al.}(2005){Anninos}, {Fragile}, \&
  {Salmonson}}]{2005ApJ...635..723A}
{Anninos}, P., {Fragile}, P.~C., \& {Salmonson}, J.~D. 2005, \apj, 635, 723,
  \dodoi{10.1086/497294}

\bibitem[{{Arfken} \& {Weber}(2005)}]{2005mmp..book.....A}
{Arfken}, G.~B., \& {Weber}, H.~J. 2005, {Mathematical methods for physicists
  6th ed.}

\bibitem[{{Ayachit}(2015)}]{paraview2015}
{Ayachit}, U. 2015, {The ParaView Guide: A Parallel Visualization Application},
  276

\bibitem[{{Bryan} {et~al.}(2014){Bryan}, {Norman}, {O'Shea}, {Abel}, {Wise},
  {Turk}, {Reynolds}, {Collins}, {Wang}, {Skillman}, {Smith}, {Harkness},
  {Bordner}, {Kim}, {Kuhlen}, {Xu}, {Goldbaum}, {Hummels}, {Kritsuk}, {Tasker},
  {Skory}, {Simpson}, {Hahn}, {Oishi}, {So}, {Zhao}, {Cen}, {Li}, \& {Enzo
  Collaboration}}]{2014ApJS..211...19B}
{Bryan}, G.~L., {Norman}, M.~L., {O'Shea}, B.~W., {et~al.} 2014, \apjs, 211,
  19, \dodoi{10.1088/0067-0049/211/2/19}

\bibitem[{{Caproni} {et~al.}(2017){Caproni}, {Amaral Lanfranchi}, {Campos
  Baio}, {Kowal}, \& {Falceta-Gon{\c c}alves}}]{2017ApJ...838...99C}
{Caproni}, A., {Amaral Lanfranchi}, G., {Campos Baio}, G.~H., {Kowal}, G., \&
  {Falceta-Gon{\c c}alves}, D. 2017, \apj, 838, 99,
  \dodoi{10.3847/1538-4357/aa6002}

\bibitem[{{Caproni} {et~al.}(2015){Caproni}, {Lanfranchi}, {da Silva}, \&
  {Falceta-Gon{\c c}alves}}]{2015ApJ...805..109C}
{Caproni}, A., {Lanfranchi}, G.~A., {da Silva}, A.~L., \& {Falceta-Gon{\c
  c}alves}, D. 2015, \apj, 805, 109, \dodoi{10.1088/0004-637X/805/2/109}

\bibitem[{{Cioffi} {et~al.}(1988){Cioffi}, {McKee}, \&
  {Bertschinger}}]{1988ApJ...334..252C}
{Cioffi}, D.~F., {McKee}, C.~F., \& {Bertschinger}, E. 1988, \apj, 334, 252,
  \dodoi{10.1086/166834}

\bibitem[{{Colella} \& {Woodward}(1984)}]{1984JCoPh..54..174C}
{Colella}, P., \& {Woodward}, P.~R. 1984, Journal of Computational Physics, 54,
  174, \dodoi{10.1016/0021-9991(84)90143-8}

\bibitem[{{Creasey} {et~al.}(2011){Creasey}, {Theuns}, {Bower}, \&
  {Lacey}}]{2011MNRAS.415.3706C}
{Creasey}, P., {Theuns}, T., {Bower}, R.~G., \& {Lacey}, C.~G. 2011, \mnras,
  415, 3706, \dodoi{10.1111/j.1365-2966.2011.19001.x}

\bibitem[{{Emerick} {et~al.}(2019){Emerick}, {Bryan}, \& {Mac
  Low}}]{2019MNRAS.482.1304E}
{Emerick}, A., {Bryan}, G.~L., \& {Mac Low}, M.-M. 2019, \mnras, 482, 1304,
  \dodoi{10.1093/mnras/sty2689}

\bibitem[{{Emerick} {et~al.}(2020){Emerick}, {Bryan}, \& {Mac
  Low}}]{2020arXiv200703702E}
---. 2020, arXiv e-prints, arXiv:2007.03702.
\newblock \doarXiv{2007.03702}

\bibitem[{{Emerick} {et~al.}(2016){Emerick}, {Mac Low}, {Grcevich}, \&
  {Gatto}}]{2016ApJ...826..148E}
{Emerick}, A., {Mac Low}, M.-M., {Grcevich}, J., \& {Gatto}, A. 2016, \apj,
  826, 148, \dodoi{10.3847/0004-637X/826/2/148}

\bibitem[{{Fragile} {et~al.}(2003){Fragile}, {Murray}, {Anninos}, \&
  {Lin}}]{2003ApJ...590..778F}
{Fragile}, P.~C., {Murray}, S.~D., {Anninos}, P., \& {Lin}, D. N.~C. 2003,
  \apj, 590, 778, \dodoi{10.1086/375183}

\bibitem[{{Fragile} {et~al.}(2004){Fragile}, {Murray}, \&
  {Lin}}]{2004ApJ...617.1077F}
{Fragile}, P.~C., {Murray}, S.~D., \& {Lin}, D. N.~C. 2004, \apj, 617, 1077,
  \dodoi{10.1086/425494}

\bibitem[{{Fryxell} {et~al.}(2000){Fryxell}, {Olson}, {Ricker}, {Timmes},
  {Zingale}, {Lamb}, {MacNeice}, {Rosner}, {Truran}, \&
  {Tufo}}]{2000ApJS..131..273F}
{Fryxell}, B., {Olson}, K., {Ricker}, P., {et~al.} 2000, \apjs, 131, 273,
  \dodoi{10.1086/317361}

\bibitem[{{Gammie} {et~al.}(2003){Gammie}, {McKinney}, \&
  {T{\'o}th}}]{2003ApJ...589..444G}
{Gammie}, C.~F., {McKinney}, J.~C., \& {T{\'o}th}, G. 2003, \apj, 589, 444,
  \dodoi{10.1086/374594}

\bibitem[{{Irwin} \& {Hatzidimitriou}(1995)}]{1995MNRAS.277.1354I}
{Irwin}, M., \& {Hatzidimitriou}, D. 1995, \mnras, 277, 1354,
  \dodoi{10.1093/mnras/277.4.1354}

\bibitem[{{Landau} \& {Lifshitz}(1987)}]{1987flme.book.....L}
{Landau}, L.~D., \& {Lifshitz}, E.~M. 1987, {Fluid Mechanics}

\bibitem[{{Lanfranchi} {et~al.}(2021){Lanfranchi}, {Hazenfratz}, {Caproni}, \&
  {Silk}}]{2021ApJ...914...32L}
{Lanfranchi}, G.~A., {Hazenfratz}, R., {Caproni}, A., \& {Silk}, J. 2021, \apj,
  914, 32, \dodoi{10.3847/1538-4357/abf6d2}

\bibitem[{{Lanfranchi} \& {Matteucci}(2004)}]{2004MNRAS.351.1338L}
{Lanfranchi}, G.~A., \& {Matteucci}, F. 2004, \mnras, 351, 1338,
  \dodoi{10.1111/j.1365-2966.2004.07877.x}

\bibitem[{{Lanfranchi} \& {Matteucci}(2007)}]{2007A&A...468..927L}
---. 2007, \aap, 468, 927, \dodoi{10.1051/0004-6361:20066576}

\bibitem[{{Lanfranchi} {et~al.}(2023){Lanfranchi}, {Soares}, \&
  {Caproni}}]{Lanfranchi2022}
{Lanfranchi}, G.~A., {Soares}, J.~F., \& {Caproni}, A. 2023, submitted

\bibitem[{{Liou}(1996)}]{1996JCoPh.129..364L}
{Liou}, M.-S. 1996, Journal of Computational Physics, 129, 364,
  \dodoi{10.1006/jcph.1996.0256}

\bibitem[{{Mac Low} \& {Ferrara}(1999)}]{1999ApJ...513..142M}
{Mac Low}, M.-M., \& {Ferrara}, A. 1999, \apj, 513, 142, \dodoi{10.1086/306832}

\bibitem[{{Marcolini} {et~al.}(2006){Marcolini}, {D'Ercole}, {Brighenti}, \&
  {Recchi}}]{2006MNRAS.371..643M}
{Marcolini}, A., {D'Ercole}, A., {Brighenti}, F., \& {Recchi}, S. 2006, \mnras,
  371, 643, \dodoi{10.1111/j.1365-2966.2006.10671.x}

\bibitem[{{Mateo}(1998)}]{1998ARA&A..36..435M}
{Mateo}, M.~L. 1998, \araa, 36, 435, \dodoi{10.1146/annurev.astro.36.1.435}

\bibitem[{{Melioli} {et~al.}(2015){Melioli}, {Brighenti}, \&
  {D'Ercole}}]{2015MNRAS.446..299M}
{Melioli}, C., {Brighenti}, F., \& {D'Ercole}, A. 2015, \mnras, 446, 299,
  \dodoi{10.1093/mnras/stu2008}

\bibitem[{{Melioli} \& {de Gouveia Dal Pino}(2015)}]{2015ApJ...812...90M}
{Melioli}, C., \& {de Gouveia Dal Pino}, E.~M. 2015, \apj, 812, 90,
  \dodoi{10.1088/0004-637X/812/2/90}

\bibitem[{{Mignone} {et~al.}(2007){Mignone}, {Bodo}, {Massaglia}, {Matsakos},
  {Tesileanu}, {Zanni}, \& {Ferrari}}]{2007ApJS..170..228M}
{Mignone}, A., {Bodo}, G., {Massaglia}, S., {et~al.} 2007, \apjs, 170, 228,
  \dodoi{10.1086/513316}

\bibitem[{{Mori} \& {Burkert}(2000)}]{2000ApJ...538..559M}
{Mori}, M., \& {Burkert}, A. 2000, \apj, 538, 559, \dodoi{10.1086/309140}

\bibitem[{{Ostriker} \& {McKee}(1988)}]{1988RvMP...60....1O}
{Ostriker}, J.~P., \& {McKee}, C.~F. 1988, Reviews of Modern Physics, 60, 1,
  \dodoi{10.1103/RevModPhys.60.1}

\bibitem[{{Raga} {et~al.}(2000){Raga}, {Navarro-Gonz{\'a}lez}, \&
  {Villagr{\'a}n-Muniz}}]{2000RMxAA..36...67R}
{Raga}, A.~C., {Navarro-Gonz{\'a}lez}, R., \& {Villagr{\'a}n-Muniz}, M. 2000,
  \rmxaa, 36, 67

\bibitem[{{Read} {et~al.}(2006){Read}, {Pontzen}, \&
  {Viel}}]{2006MNRAS.371..885R}
{Read}, J.~I., {Pontzen}, A.~P., \& {Viel}, M. 2006, \mnras, 371, 885,
  \dodoi{10.1111/j.1365-2966.2006.10720.x}

\bibitem[{{Recchi}(2014)}]{2014AdAst2014E...4R}
{Recchi}, S. 2014, Advances in Astronomy, 2014, 750754,
  \dodoi{10.1155/2014/750754}

\bibitem[{{Revaz} {et~al.}(2009){Revaz}, {Jablonka}, {Sawala}, {Hill},
  {Letarte}, {Irwin}, {Battaglia}, {Helmi}, {Shetrone}, {Tolstoy}, \&
  {Venn}}]{2009A&A...501..189R}
{Revaz}, Y., {Jablonka}, P., {Sawala}, T., {et~al.} 2009, \aap, 501, 189,
  \dodoi{10.1051/0004-6361/200911734}

\bibitem[{{Ruiz} {et~al.}(2013){Ruiz}, {Falceta-Gon{\c{c}}alves}, {Lanfranchi},
  \& {Caproni}}]{2013MNRAS.429.1437R}
{Ruiz}, L.~O., {Falceta-Gon{\c{c}}alves}, D., {Lanfranchi}, G.~A., \&
  {Caproni}, A. 2013, \mnras, 429, 1437, \dodoi{10.1093/mnras/sts425}

\bibitem[{{Silich} \& {Tenorio-Tagle}(1998)}]{1998MNRAS.299..249S}
{Silich}, S.~A., \& {Tenorio-Tagle}, G. 1998, \mnras, 299, 249,
  \dodoi{10.1046/j.1365-8711.1998.01765.x}

\bibitem[{{Simpson} {et~al.}(2015){Simpson}, {Bryan}, {Hummels}, \&
  {Ostriker}}]{2015ApJ...809...69S}
{Simpson}, C.~M., {Bryan}, G.~L., {Hummels}, C., \& {Ostriker}, J.~P. 2015,
  \apj, 809, 69, \dodoi{10.1088/0004-637X/809/1/69}

\bibitem[{{Springel}(2005)}]{2005MNRAS.364.1105S}
{Springel}, V. 2005, \mnras, 364, 1105,
  \dodoi{10.1111/j.1365-2966.2005.09655.x}

\bibitem[{{Stinson} {et~al.}(2007){Stinson}, {Dalcanton}, {Quinn}, {Kaufmann},
  \& {Wadsley}}]{2007ApJ...667..170S}
{Stinson}, G.~S., {Dalcanton}, J.~J., {Quinn}, T., {Kaufmann}, T., \&
  {Wadsley}, J. 2007, \apj, 667, 170, \dodoi{10.1086/520504}

\bibitem[{{Stone} \& {Norman}(1992)}]{1992ApJS...80..753S}
{Stone}, J.~M., \& {Norman}, M.~L. 1992, \apjs, 80, 753, \dodoi{10.1086/191680}

\bibitem[{{Teyssier}(2002)}]{2002A&A...385..337T}
{Teyssier}, R. 2002, \aap, 385, 337, \dodoi{10.1051/0004-6361:20011817}

\bibitem[{Toro(2009)}]{toro2009riemann}
Toro, E. 2009, Riemann Solvers and Numerical Methods for Fluid Dynamics: A
  Practical Introduction (Springer Berlin Heidelberg).
\newblock \url{https://books.google.com.br/books?id=SqEjX0um8o0C}

\bibitem[{{Wada} \& {Venkatesan}(2003)}]{2003ApJ...591...38W}
{Wada}, K., \& {Venkatesan}, A. 2003, \apj, 591, 38, \dodoi{10.1086/375335}

\bibitem[{{Zingale} {et~al.}(2002){Zingale}, {Dursi}, {ZuHone}, {Calder},
  {Fryxell}, {Plewa}, {Truran}, {Caceres}, {Olson}, {Ricker}, {Riley},
  {Rosner}, {Siegel}, {Timmes}, \& {Vladimirova}}]{2002ApJS..143..539Z}
{Zingale}, M., {Dursi}, L.~J., {ZuHone}, J., {et~al.} 2002, \apjs, 143, 539,
  \dodoi{10.1086/342754}

\end{thebibliography}
\bibliographystyle{aasjournal}



\end{document}